\documentclass[preprint]{aastex}
\usepackage{emulateapj5}
\usepackage{epsfig}

\begin{document}

\title{Chaos caused by resonance overlap in the solar neighborhood  --
Spiral structure at the Bar's 2:1 outer Lindblad resonance 
}

\author{A.~C.\ Quillen
}
\affil{Department of Physics and Astronomy, University of Rochester, Rochester, NY 14627; aquillen@pas.rochester.edu}
\affil{Visitor, Physics Department, Technion, Israel Institute of Technology}

\begin{abstract}
We consider the nature of orbits
near the solar neighborhood which are perturbed by 
local spiral arms and the Milky Way bar.
We present a simplified Hamiltonian model which includes resonant terms
from both types of perturbations and is similar to the forced pendulum.  
Via numerical integration of this model 
we construct Poincare maps to illustrate the
nature and stability of the phase space. 
We find that resonance overlap is most likely
to cause widespread chaos when the pattern
of the spiral structure puts the solar
neighborhood near the 2:1 inner Lindblad resonance 
(ILR) in the case of a 2-armed pattern, 
or near the 4:1 ILR in the case of a 4-armed pattern.
When this happens both the quasiperiodic orbits which support
the spiral structure and those that oscillate with the bar
are disrupted near the bar's 2:1 outer Lindblad resonance (OLR).
Consequently the pattern speed of spiral structure
which passes through the OLR must be faster than
$\sim 0.45$ times the solar neighborhood angular rotation rate
if it is 2-armed or faster than $0.75$ times this value
if it is 4-armed.
Alternatively the OLR may form a boundary between spiral
modes at different pattern speeds.
In all cases we find that spiral structure is disrupted by the 
OLR over a narrow range of radius and the extent of the 
orbits aligned perpendicular
to the bar at the OLR is limited by the spiral perturbations.
We find that the boundary at an asymmetric drift velocity
of $v\sim -30$ km/s of the $u$-anomaly in the velocity distribution
in the solar neighborhood is due to the abrupt bifurcation
of the orbit families associated with the OLR.
The upper boundary at $v\sim -60$ km/s is truncated by the spiral structure.
The radial velocity width of the anomaly is probably bounded by 
chaotic regions which surround the family of quasiperiodic  orbits
oriented perpendicular to the bar.

\end{abstract}


\section{Introduction}

It has been established, beyond doubt, that the Milky is barred
as are many nearby galaxies.
Evidence for the bar comes from asymmetry
in the bulge surface photometry,
star counts and interpretation of the gas kinematics (for a recent
review see \citealt{gerhard}).
The most common values suggested from these observations
are a corotation radius of 3.5-5 kpc and a major axis
in-plane angle with respect to the Galactic Center
of $\phi_{bar} \sim 15-45^\circ$.

The solar neighborhood local velocity distribution
is expected to contain structure caused by large scale
nearby gravitational perturbations.  
While structure at low asymmetric drift is likely to be
affected by spiral structure and disrupted stellar
clusters (e.g., \citealt{skuljan}), 
the Hercules stream at a mean heliocentric asymmetric
drift velocity of $v \sim 45$km/s is suspected to be related
to the 2:1 outer Lindblad resonance (OLR) 
with the Galactic bar \citep{dehnen99,fux,raboud}.
This stream is also known as the $u$-anomaly. 
Due to the older and rather high metallicity
stars in it, the stream probably has a stable kinematic origin 
\citep{raboud,dehnen98}.
The local velocity distribution as constructed from
Hipparcos catalog most clearly shows
this stream as a strong and separate feature \citet{dehnen98};
in fact there is a saddle point in the distribution between the stream
and the bulk of the stars at $v\sim 30$ km/s (see Figures
in \citealt{dehnen98,fux}).

Linear theory predicts that the orientation of orbits
will shift across the OLR, from oriented
along the bar major axis outside the OLR to perpendicular
to the bar within it \citep{B+T}.  Near the peak of the resonance
both types of orbits can exist \citep{cont75, weinberg}.
\citet{dehnen20} showed using a backwards integrating 
bar growth model that stars were likely to be captured
into resonant islands associated with these 2 major orbit
families.  However the final
velocity distribution is strongly dependent
upon the timescale over which the bar grows as well
as the assumed initial velocity distribution.
\citet{fux} considered the stability of orbits 
and suggested that chaotic regions associated with resonances were
likely to cause over-densities in some regions
of phase space and under-densities in others. 
He illustrated with N-body simulations that the resulting
stellar distribution could correspond to that observed.

In some sense it is surprising that the OLR with
the bar provides such a good explanation
for the $u$-anomaly because locally both gaseous and stellar
tracers primarily show spiral structure near the solar
circle (e.g., \citealt{vallee,drimmel}).
Previous works have not considered the more complicated
problem caused by the forces due to local spiral structure
in addition to those caused by the bar.

In this paper we consider the dynamics of stars that
are affected by perturbations from both spiral structure
and the Milky Way bar.  
We construct a simple Hamiltonian
model for the dynamics which contains resonant
terms at both the bar pattern speed and 
a slower pattern speed from local spiral structure.
We find that the Hamiltonian can exhibit resonance overlap
and so widescale chaos.
Consequently we address the issue of orbit stability 
by computing area preserving or Poincare maps 
for different values of the spiral pattern speed.




\section{Hamiltonian Formalism for the Kinematics}


As shown by \citet{cont75,cont88},
the dynamics of stars confined to the galactic plane
moving in a smooth Galactic potential 
lacking non-axisymmetric perturbations 
is described by a Hamiltonian 
which can be written in a third order post epicyclic approximation as
\begin{equation}
\label{H0}
H_0(I_1,\theta_1;I_2,\theta_2) 
= h + \omega_1 I_1 + \omega_2 I_2 + a I_1^2 + 2 b I_1 I_2 + c I_2^2 ...
\end{equation}
The high order of the expansion is required to exhibit
the bifurcation of the major resonances.
The action variables $I_1 = {1\over 2 \pi} \int \dot r dr$ and $I_2 = J_0 - J_c$ 
are integrals of motion when the Hamiltonian is unperturbed.
$J_0$ is the particle's angular momentum and $J_c$ the angular
momentum of a particle in a circular orbit at a radius $r_c$ 
which is the radius of a circular orbit with energy $h$.
The frequencies and constants in the above Hamiltonian are evaluated at this 
radius.
We can either to choose to work in a frame rotating with a perturbation pattern or
an inertial one.
\cite{cont75} worked in a frame rotating at the bar or spiral pattern speed, $\Omega_s$,
so $h \equiv {J_c^2 \over 2 r_c^2} + V_0(r_c) -\Omega_s J_c$ and 
$\omega_2 = \Omega_c - \Omega_s$ where $\Omega_c$ is the angular rotation rate
of a circular orbit. 
Since we will consider more than one perturbation  and each perturbation
will have a different pattern speed, in this paper 
we choose to work in an inertial frame.
Consequently
$h \equiv {J_c^2 \over 2 r_c^2} + V_0(r_c)$ and  $\omega_2 = \Omega_c$.
In either frame, $J_c = r_c^2 \Omega_c$ 
is the angular momentum and $V_0(r)$ the axisymmetric component
of the potential.
The frequency $\omega_1 = \kappa_c$ where $\kappa_c$ is the  epicyclic frequency
which is
evaluated at $r_c$.   Unfortunately the value of $r_c$ depends on 
$\Omega_s$ so transferring coordinate systems between an inertial 
and a rotating one is not trivial.

The constants $a,b,c$ are given in the appendix of \cite{cont75}.
When the rotation curve is flat $\kappa_c = \sqrt{2} \Omega_c$,
$a =-{0.92 \over r_c^2}$,
$b = -{1 \over \sqrt{2} r_c^2}$, and $c= -{1 \over 2 r_c^2}$. 

The radius $r$, the angle $\theta$ and their time derivatives 
are related to the action angle variables
to first order in $I_1^{1\over 2}$
(from the appendix of \citealt{cont75})
\begin{eqnarray}
r-r_c \approx \left({2 I_1\over \kappa_c}\right)^{1\over 2}\cos{\theta_1} \\
{dr\over dt} \approx - (2 I_1 \kappa_c)^{1\over 2} \sin{\theta_1}\\
\theta \approx \theta_2 - {2 \Omega_c \over r_c \kappa_c} \left({2 I_1 \over \kappa_c }\right)^{1\over 2} 
\sin{\theta_1}  \\
r^2{d\theta\over dt} \approx I_2 + r_c^2 \Omega_c
\end{eqnarray}
where $\theta_1$ is the epicyclic angle and $\theta_2$ is
the azimuth of the epicyclic center.  $\theta_1$ and $\theta_2$
are the angle variables conjugate to $I_1$ and $I_2$.

To consider the affect of non-axisymmetric perturbations in the gravitational
potential caused by a bar or spiral structure we must estimate
the strength of these perturbations in terms of the action
angle variables described above.
We concentrate on the Lindblad resonances for a number of reasons.
They are first order in the amplitude of the perturbation 
and in ${I_1}^{1\over 2}$
and so likely to be strong.
The 2:1 OLR from the Galactic bar is near the location of the solar neighborhood.
Stars on orbits influenced by the ILRs (inner Lindblad resonances) 
are required to self consistently support spiral structure 
(e.g., \citealt{cont88}).

\subsection{The perturbation from the Milky Way Bar}

For a bar like perturbation we assume that the non-axisymmetric
component of the gravitational potential depends on radius and 
can be expanded in Fourier components 
\begin{equation}
V_{1}(r,\theta) = \sum_m B_m(r) \cos[m (\theta - \Omega_b t) ]
\end{equation}
where $\Omega_b$ is the angular rotation rate of the bar.
In action angle variables we can write the potential perturbations 
to first order in $I_1^{1/2}$ and the strength of
the perturbation (following expressions given by \citealt{cont88})
\begin{eqnarray}
V_{1,m}&(I_1,\theta_1;I_2,\theta_2) = 
\left({2 I_1 \over \kappa_c}\right)^{1\over 2}  \beta_m
\times  ~~~~~~~~~~~~~~~~~~~~  \\
& \left[
{ \cos(\theta_1 + m (\theta_2-\Omega_b t))  + \cos(\theta_1 - m (\theta_2 -\Omega_b t))}
\right] \nonumber 
\end{eqnarray}
where the first and second angular
terms correspond to the $m:1$ outer and inner Lindblad resonances 
respectively.  Here
\begin{equation}
\beta_m = 
\left[{{1\over 2} B_m'  +  {m \Omega_c \over r_c \kappa_c }B_m } \right]
\label{barper}
\end{equation}
where $B_m$ and $B_m'$ are evaluated at $r_c$. 

The solar neighborhood is well past the end of the Galactic bar
so we can approximate the potential perturbation as a
quadrupolar term in the  gravitational potential.
\begin{equation}
B_2(r) \approx - a_b \left({r_b\over r}\right)^3
\end{equation}
where $r_b$ is the radius at which the bar ends, 
$r_b \approx  0.45 r_0$,
as found from IR photometry (e.g., \citealt{dwek}) where
$r_0$ is the radius of the solar circle from the Galactic Center. 
The pattern speed of the bar is constrained by 
to be about 1.85 times the solar neighborhood value of
the angular rotation rate, $\Omega_0$ \citep{dehnen99}.
\citet{fux} considered bars 
about twice as strong as those considered by \citet{dehnen99}.

\cite{dehnen20} estimates
$\alpha =  {3 a_b \over v_c^2} \left({r_b \over r_c}\right)^3  \approx 0.01$,
so at the solar circle 
${a_b \over v_c^2} \sim 0.036.$
From equation (\ref{barper}) and assuming that the rotation
curve is flat we find that  
$
\beta_2 \sim  0.086 {a_b \over r_c }  \left({r_b\over r_c}\right)^3 
$
so at the solar neighborhood we estimate 
\begin{equation}
\beta_2  \sim  0.003  {v_c^2 \over r_c}
\end{equation}
where $v_c$ is the velocity of a star in a circular orbit.

\subsection{The perturbation from local spiral structure}

For spiral structure we assume that the radial variations
depend on angle and the amplitude is nearly constant with radius.
\begin{equation}
V_{1,m}(r,\theta) = {\rm Re}\left[ 
\sum_{m} A_m
\exp \left[i (k_m r - m(\theta-\Omega_s t) + \alpha_m )\right]
\right]
\end{equation}
Here $m$ refers to the number of arms, $\Omega_s$ to the spiral
pattern speed, $k_m$ to the wavenumber and $\alpha_m$ to an angular
offset.

\cite{cont75} showed to first order in $I_1^{1/2}$ the potential
perturbation
\begin{eqnarray}
&V_{1,m}(I_1,\theta_1;I_2,\theta_2) = 
\left({2 I_1 \over \kappa_c}\right)^{1 \over 2}  \epsilon_m \times ~~~~~~~~~~~~~~~~~ \\
&\left[
 \cos(\theta_1 - m(\theta_2 - \Omega_s t + \gamma_{m+})) +  
 \cos(\theta_1 + m(\theta_2 - \Omega_s t+ \gamma_{m-}))
\right] \nonumber 
\end{eqnarray}
where 
\begin{eqnarray}
\label{epsilon_m}
 \epsilon_{m} &=& {A_m  k_m \over 2} \sqrt{ 
 1 +  \left({2 m \Omega_c \over k_m r_c \omega_1}\right)^2
}
\end{eqnarray}
for ${ A_m' \over k A_m} \ll 1$.
The angular offsets $\gamma_{m\pm}$ depend on $\alpha_m$ and the wavenumber
(see \citealt{cont75} for expressions).

Tracers of spiral structure in the solar neighborhood
suggest that the spiral structure
is quite tightly wound (e.g., \citealt{vallee}).
In the WKB approximation 
\begin{equation}
A_m \sim {-2 G \Sigma_0 A \over| k_m|}
\end{equation}
where $S_m$ is the amplitude of the spiral surface density variations 
divided by the mean surface density \citep{B+T}.
For a tightly wound structure  we expect that 
 $ \epsilon_m \sim {A_m  k_m \over 2} $ 
(taking the largest term in equation \ref{epsilon_m}).
The mean surface density of disk mass in the solar 
solar neighborhood is $\Sigma_0 \sim 50 M_\odot {\rm pc}^{-2}$ \citep{holmberg},
and we use $v_c \sim 200 $km/s.  
A two armed spiral is seen in the near infrared COBE/DIRBE data
with pitch angle in the range $p\sim 15.5^\circ$~--~$19^\circ$ \citep{drimmel},
giving $k_m r_0 = m \cot{p}$ in the range 5.8~--~7.2.
The amplitude of the spiral structure could be as large as
$S_m\sim 0.5-1$ \citep{drimmel}.
We estimate for the solar neighborhood
\begin{eqnarray}
 A_m  \sim  & -0.013 v_c^2 S_m
 \left({\Sigma_0 \over 50 M_\odot {\rm pc}^{-2} }\right)
 \left({v_c \over 200 {\rm km~s}^{-1}}\right)^{2}
 \left({r_c \over 8 {\rm kpc} }\right)
 \left({7 \over k_m r_c }\right) \nonumber \\
\epsilon_m   \sim & -0.025  {v_c^2 \over r_c} S_m
 \left({\Sigma_0 \over 50 M_\odot {\rm pc}^{-2} }\right)
 \left({v_c \over 200 {\rm km~s}^{-1}}\right)^{2}
 \left({r_c \over 8 {\rm kpc} }\right).  ~~~~~~~
\end{eqnarray}
The expression for $\epsilon_m$ is independent of $k_m$ and primarily
depends on the amplitude of the spiral density variation.



\section{Hamiltonian with perturbations from both spiral structure and the Milky Way Bar}

We now take the perturbations due to the bar and spiral arms
and add them to the unperturbed Hamiltonian.
We have estimated the size of the coefficients 
$\beta$ and $\epsilon$ for these perturbations in the previous section.

We expect that the 2:1 OLR with the Galactic bar and the 2:1 or 4:1 
ILR with local spiral
structure will be the strongest resonances near the solar neighborhood.
So ignoring all other resonant terms we can simplify the problem to
\begin{eqnarray}
\label{barspi}
H &= a I_1^2 + \kappa I_1 +\Omega I_2 
        + \beta_2    ({2 I_1\over \kappa})^{1 \over 2} \cos[ \theta_1 + 2(\theta_2 - \Omega_b t)] 
\nonumber \\
& + \epsilon_m ({2 I_1\over \kappa})^{1 \over 2} \cos[ \theta_1 - m(\theta_2 - \Omega_s t) - \gamma]   + ....
\end{eqnarray}
where $m=2$ for the 2:1 ILR and $m=4$ for the 4:1 ILR.
We have assumed that $I_2$ is small and so have dropped 
the terms $b I_1 I_2$ and $c I_2^2$ from the Hamiltonian.

We do a canonical transformation with generating function
\begin{equation}
F_2(J_1,J_2,\theta_1,\theta_2) = 
J_1 (\theta_1 + 2( \theta_2 - \Omega_b t)) + J_2 \theta_2
\end{equation}
obtaining a resonant angle and new momenta
\begin{eqnarray}
\phi = \theta_1 + 2(\theta_2 - \Omega_b t) \nonumber \\
I_2 = 2 J_1 + J_2. 
\end{eqnarray}
$I_1=J_1$ and $\theta_2$ remain unchanged.
Our Hamiltonian becomes
\begin{eqnarray}
H(J_1,\phi; J_2,\theta_2) &= a J_1^2 + \delta J_1 + \Omega J_2  
     + \beta_2    ({2 J_1\over \kappa})^{1 \over 2} \cos(\phi) \nonumber \\
& + \epsilon_2 ({2 J_1\over \kappa})^{1 \over 2} \cos(\phi + \alpha  - \gamma) 
\end{eqnarray}
where  
\begin{eqnarray}
\delta = \kappa + 2 (\Omega - \Omega_b) \\
\alpha =    (m \Omega_s + 2\Omega_b)t - ( m+2) \theta_2.
\end{eqnarray}
We can approximate $\theta_2 \approx \Omega t$ so that
$\alpha \approx  \nu t $ where 
\begin{equation}
\nu  =   m \Omega_s + 2\Omega_b - (2+m) \Omega.
\end{equation}
We rewrite our Hamiltonian in a simpler form
\begin{eqnarray}
H(J_1,\phi) &\approx a J_1^2 + \delta J_1 
     - \beta_2    ({2 J_1\over \kappa})^{1 \over 2} \cos(\phi) \nonumber \\
& - \epsilon_2 ({2 J_1\over \kappa})^{1 \over 2} \cos(\phi + \nu t  -\gamma) .
\end{eqnarray}
We rescale by dividing by $a$, put lengths in units of $r_c$
and time in units of ${1\over \kappa}$ so
that $J_1 ={j_1 \kappa r_c^2}$.
Since $a<0$ this has the effect of reversing the sign of $t$.
\begin{equation}
\label{ham_integrate}
h(j_1,\phi) \approx j_1^2 + \bar\delta j_1 
     - \bar \beta_2    j_1^{1 \over 2} \cos(\phi) \nonumber 
 - \bar \epsilon_2 j_1^{1 \over 2} \cos(\phi - \bar \nu t  - \gamma) 
\end{equation}
where the unitless coefficients
\begin{eqnarray}
\bar\delta   &=& {\kappa + 2 (\Omega - \Omega_b) \over a r_c^2  \kappa} \\
\bar\beta    &=& {-\beta_2 \sqrt{2} \over a r_c^3 \kappa^2} \\
\bar\epsilon &=& {-\epsilon_2 \sqrt{2} \over a r_c^3 \kappa^2} \\
\bar\nu      &=& {m \Omega_s + 2\Omega_b - (2+m) \Omega\over \kappa}.
\end{eqnarray}

If we set $\bar\epsilon=0$ then
$h = j_1^2 + \bar\delta j_1 - \bar\beta j^{1\over2} \cos{\phi}$.
This Hamiltonian
is in the same form as $e-e$ resonances discussed in the context
of solar system orbital resonances 
\citep{SolarSystemDynamics,borderies}.
As illustrated by \citet{SolarSystemDynamics,borderies}
the resonance bifurcates at a  critical value of a parameter
which depends on  $\bar\delta$
and $\bar\beta$ (see Figure 1).  
The resonance has three fixed points when 
\begin{equation}
{2 \bar\delta \over 3 \bar\beta^{2\over 3}} < -1.
\end{equation}
One of the fixed points is unstable, the other two 
correspond to two resonant stable islands of libration at $\phi = 0,\pi$.
Only one fixed point exists when the above inequality
does not hold, and its location
is determined by the sign of $\bar\beta$. When $\bar\beta>0$
this fixed point is at $\phi=0$, otherwise at $\phi=\pi$.

When $\bar\beta=0$ the Hamiltonian
can be put in the form
$h = j_1^2 + (\bar\delta -\bar\nu)j_1 - \bar\epsilon j^{1\over2} \cos{\phi'}$
with a suitable canonical transformation.
In this case the bifurcation happens when 
\begin{equation}
{2 (\bar\delta-\bar\nu ) \over 3 \bar\epsilon^{2\over 3}} < -1.
\end{equation}

The distinction between regions with two stable fixed points and 
with one fixed point is important when resonances
overlap.   For this type of Hamiltonian 
a separatrix exists only when there are two stable fixed points.
Because the period of the orbits becomes infinite at the separatrix,
additional perturbations are most likely to cause instability and a chaotic
region near the original location of the separatrix.
When we have two resonant terms
in the Hamiltonian and the resonances overlap, large regions of
phase space will become strongly chaotic only when 
at least one of the resonances contains a separatrix.

\smallskip
{
\centering
\includegraphics[angle=0,width=2.5in]{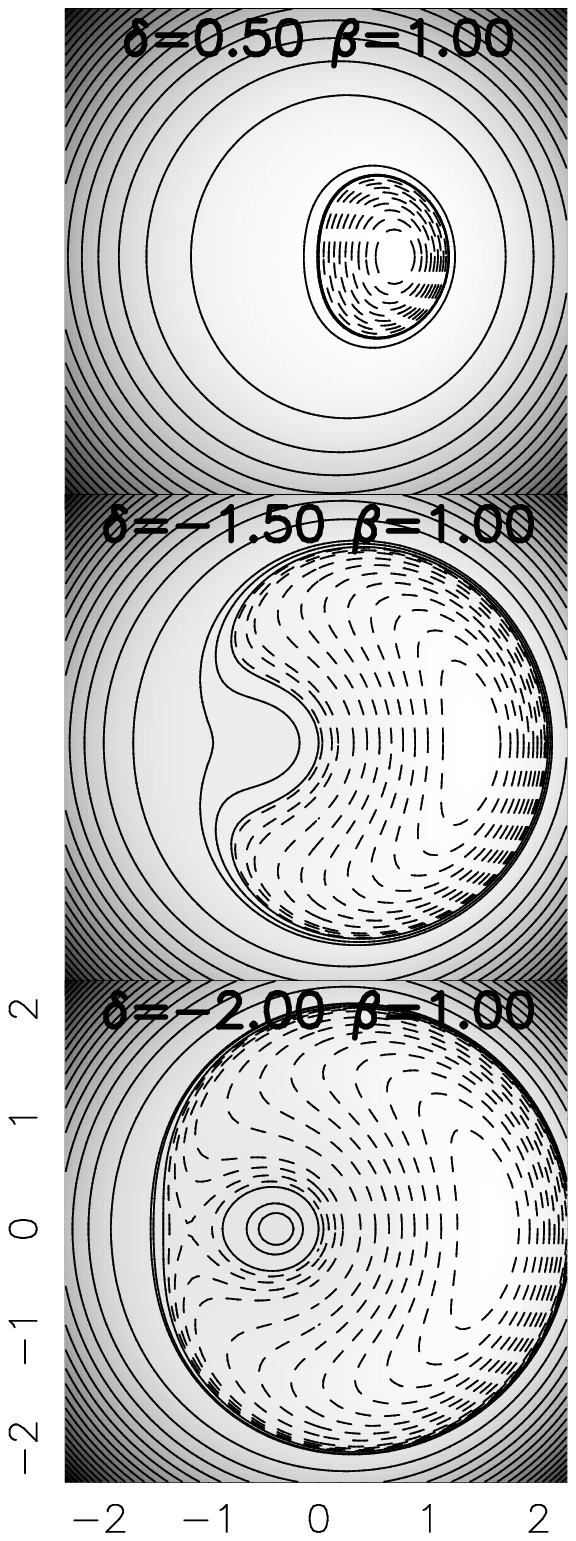}
}
\begin{quote}
\baselineskip3pt
\footnotesize
Fig.~1.--- We plot level contours of the Hamiltonian 
$H(j,\phi) = j^2 + j \bar\delta - \bar \beta j^{1/2} \cos{\phi}$.
Dashed contours are negative.
The axes are $x,y$ are defined by Eqn(\ref{coord}).
The critical value for the resonance to bifurcate
happens in the middle panel where 
${2 \bar\delta \over 3 \bar\beta^{2\over 3}} = -1$.
The top of the panel contains a fixed point which would correspond to  a
closed or periodic orbit aligned with the bar so that $\phi=0$.  This is the
situation outside the OLR.
Inside the OLR the resonance bifurcates, 
and both periodic orbit families are present.  This situation
is shown in the bottom panel.  The fixed points at $\phi=\pi$
correspond to periodic orbits oriented perpendicular to the bar.
Only when ${2 \bar\delta \over 3 \bar\beta^{2\over 3}} < -1$
is there a separatrix.  Additional perturbations are most likely
to cause instability when there is a separatrix.
\end{quote}

We now consider the orientations of the orbits
and signs of the individual terms.
Fixed points in our simple one dimensional system correspond
to closed or periodic orbits in the full 2 dimensional system.
Assuming that $\Omega_b$ is 1.85  times the angular rotation
rate at the solar circle \citep{dehnen99} and a flat rotation curve we estimate
$\bar\delta \sim 0.2$ at the solar circle.
This drops to zero as we  approach the 2:1 OLR at a Galactic
centric radius of $r\approx 0.9 r_0$.
Because $a\sim -1$, $\bar\beta \sim  \beta_2 r_c /v_c^2 $ and
$\bar\epsilon \sim  \epsilon r_c /v_c^2$.
For radii outside the OLR, $\bar\delta>0$ and we expect
only one fixed point or periodic orbit.  
Because $\bar\beta >0$ the quasiperiodic orbits at the solar
circle, outside the OLR, will be 
aligned with the bar.  They are referred to 
as the $x_1(1)$ orbits (e.g., see \citealt{fux,dehnen20}).
Inside the OLR the resonance bifurcates and 
two families of periodic orbits (fixed points) exist,
the $x_1(1)$ and $x_1(2)$ families. 
The $x_1(2)$ family is aligned perpendicular to the bar
and found at $\phi = \pi$.
We expect large scale chaos can occur when  there is a separatrix,
$\bar\delta <0$, and inside the radius of the OLR.

$\Omega_s$ is $\sim 0.3$ times the angular rotation
rate at the solar circle when $r_0$ 
is just outside the 2:1 ILR  
and 0.6 if $r_0$ is just within the 4:1 ILR.
$\gamma$ is related to the 
phase of the potential for the spiral pattern.
and is difficult to constrain from observations.
For spiral structure at our location
in the solar neighborhood we should be outside the m:1
IRL of the pattern so that $\bar\delta - \bar\nu <0$.
If we aren't outside the resonance, the stellar
orbits will not support the spiral structure, and
will negate it instead (e.g., \citealt{cont88}).  
We expect $\bar\epsilon <0$
for periodic orbits with $\phi = \gamma$ to be 
in phase with the spiral arm.  By in phase we mean that
the orbit will be aligned with radial maxima on the same axis 
as the density maxima.

\subsection{Generating Poincare maps}

Previous works addressing the stability of the orbits
in the solar neighborhood have computed Liapunov exponents,
studied N-body simulations \citep{fux} or carried out
backwards integration schemes \citep{dehnen20}.
Here we adopt the approach of \citet{fux}
and strive to identify regions which regions of phase space that can 
keep stars for long periods of time.
However as we explain below we do this by mapping
the nature of phase space instead of computing Liapunov exponents.

The Hamiltonian (Eqn.\ref{ham_integrate}) contains two main
resonant terms separated by a frequency, $\nu$.  
This is similar to the forced pendulum.
When the two resonances overlap,
a chaotic region can be generated at   
the separatrix of one of the resonances.
This zone has a Liapunov time $\sim{2 \pi \over \nu}$
\citep{holman}.
Outside the main chaotic zone we expect quasiperiodic islands.
This picture is qualitatively different
than that explored by \citet{fux} who illustrated narrow bands of
unstable regions, each with different Liapunov timescales.
We expect instead large bands of chaos described by 
one Liapunov time which are surrounding stable islands.
Consequently we do not
compute Liapunov times for a range of initial conditions
but instead map the structure of the phase space itself.

To investigate the stability of the system containing perturbations
from both the bar and spiral structure we integrate
Eqn(\ref{ham_integrate}) numerically.
This system is time dependent so the Hamiltonian itself is
not conserved.  However if we plot points every timestep
$ P_\nu = {2 \pi \over \nu}$,
we derive an area preserving or Poincare map.
This procedure generates maps which are like surfaces of section,
and these we can use to study the stability of the orbits.
In such a map orbits are either area filling or curved loops.
We denote area filling orbits as chaotic (in the sense
that orbits diverge exponentially) 
and the curved orbits as quasiperiodic.

This procedure is excellent for addressing the question
of orbit stability and identifying the regions accessible 
to individual orbits. 
But because we plot every $P_\nu$ we cannot
tell if an orbit is librating around $\phi$ or around an angle
$\phi -\nu t$.  This makes it difficult to determine
whether fixed points are associated with (or supporting) 
the bar or the spiral pattern.
However we can qualitatively deal with this problem by
changing the orientation of the spiral perturbation.
We do this by adjusting $\gamma$.
We set $\gamma = {\pi \over 2}$ so that orbits 
in phase with the spiral pattern 
are located at an angle of $\phi \sim {\pi\over 2}$.
Orbits perpendicular to the spiral pattern
are located at an angle of $-{\pi\over 2}$.
In comparison, 
orbits associated with the bar's OLR librate about $\phi = 0$ or $\pi$.

As is commonly done, (e.g., \citealt{SolarSystemDynamics,borderies}),
we plot all figures in this paper in a coordinate system with 
\begin{eqnarray}
\label{coord}
x &=& \sqrt{2 j_1} \cos{\phi} \\
y &=& \sqrt{2 j_1} \sin{\phi}. \nonumber
\end{eqnarray}
With respect to our action variable, $j_1 ={s^2/ 2}$
where $s$ is the radial distance on these plots.
Since we defined 
$J_1 = j_1 \kappa r_c^2$,
the epicyclic amplitude is approximately the same
as $s$, the radial distance in these plots.  
In other words $s$ gives the maximum difference
between the radial position of a particle and $r_c$.

For each series of integrations (Figures 2--4) 
we assume a value for $\Omega_s$ and $\Omega_b$ (in units
of the solar neighborhood angular rotation rate $\Omega_0$, 
and $\bar\epsilon$ and $\bar\beta$ which are directly
estimated from the strengths of the spiral structure and bar
(see previous sections).   
We define the radial offset, $dr$, from the radius
of the sun as
\begin{equation}
dr = {r_0- r_c \over r_c}.
\end{equation}
so that 
${\Omega_0 \over \Omega} = {r_c \over r_0} = 1 + dr$.
$\bar\delta$ and $\bar\nu$ for each integration 
are calculated from $\Omega_s$ and $\Omega_b$ 
assuming a flat rotation curve and for
a range of $dr$.  In the figures the different
panels correspond to different values of $dr$.

Considering the bar only,
as we vary $dr$ and so $\delta$, phase space changes
in appearance from that shown at the top of Figure 1
to that in the bottom of this figure as we pass through
the OLR.  
Considering the spiral arms only, phase space should
look like the bottom of Figure 1 but rotated by 90 degrees since
we choose $\gamma= \pi/2$.  
When $\delta<0$, the bar OLR gains a separatrix which
is likely to become unstable when perturbed by the spiral arms.

In Figure 2 we show the result of integrating Eqn.(\ref{ham_integrate})
for a 2-armed pattern near the 4:1 ILR of the spiral pattern
for moderate spiral and bar strength.    
For each of 30 particles chosen with different initial conditions,
200 timesteps are plotted with $dt =  2\pi/\nu$. 
For the panel on the top left, corresponding
to the Galactic radius of the sun, the orbits librate
about $\phi=\pi/2$ and support the spiral structure.
At $dr \sim-0.90$, orbits become are aligned with the bar
($x_1(1)$ type orbits) and librate about $\phi=0$. 
At $dr=-0.095$ the $x_1(2)$ orbits appear.  
These librate about $\phi=-\pi$ and are oriented perpendicular
to the bar.   
As $dr$ decreases further the orbits
librate closer and closer to $\phi=\pi/2$ and so support
the spiral structure and are no longer
aligned perpendicular to the bar.  
For the intermediate region between $dr=-0.10$ and $-0.175$
the quasiperiodic orbits are likely to be oscillating with both
the bar and the spiral structure.  

Just past the OLR 
(the center of the resonance is near $\delta =0$, but a separatrix
appears only at negative values for $\delta$),
we see in Figure 2 that the quasiperiodic regions
are bounded by thick bands of area filling or chaotic orbits.
The spiral structure is disrupted (unsupported)
between $dr = -0.075$ and $-0.115$.   
Past $dr=-0.175$ the $x_1(2)$ orbits are 
limited by the influence of the spiral structure.
In short the spiral structure limits the extent of the orbits
perpendicular to the bar OLR and the bar OLR 
disrupts the spiral structure over a narrow range of radius.

\subsection{On the $u,v$ plane}

To see if our maps correspond to what is 
observed in the local velocity distribution 
we must consider what values of our action
angle variables correspond to the $u,v$ velocities
that are used to measure stars in the solar neighborhood.

The solar neighborhood velocity distribution is typically
plotted with respect to the asymmetric drift velocity (azimuthal),
$v$, and the radial velocity, $u$, where $u>0$ corresponds
to velocities toward the Galactic center.
Orbits at lower $v$ values have lower $r_c$ values and 
are expected to oscillate about galactic
radii that are smaller than $r_0$, the radius of the solar circle.
We now estimate what values of $dr$ correspond to
the center and boundary of the Hercules stream or $u$-anomaly.

If we assume a flat rotation curve, the Hamiltonian lacking non-axisymmetric 
perturbations
\begin{equation}
{H\over v_c^2} =  {u^2\over 2 v_c^2} + {(v_c+v)^2\over 2 v_c^2} + \ln{r/r_0}.
\end{equation}
For a circular orbit of radius $r_c$ both $u,v=0$ and
\begin{equation}
{H\over v_c^2}  = 1.0 + \ln{r_c/r_0}.
\end{equation}
We set $r=r_0$ at the solar neighborhood, 
equate the two expressions 
and solve for $r_c$ as a function of the asymmetric drift velocity $v$,
finding   $\ln{r_c \over r_0} = {v\over v_c} + {u^2 + v^2\over 2 v_c^2}$.
We expand $\ln{r_c \over r_0}$ and obtain 
\begin{equation}
dr = {v\over v_c} (1 + {v\over2 v_c }) + {u^2 \over v_c^2}. 
\end{equation}
Assuming $u=35$km/s at the center of the stream and $v_c = 200$km/s 
\begin{eqnarray}
dr \sim -0.14 ~~  & &~~  {\rm for} ~ v \sim -30 {\rm km~s^{-1}} \\
dr \sim -0.18 ~~  & &~~  {\rm for} ~ v \sim -45 {\rm km~s^{-1}}
\nonumber 
\end{eqnarray}
where $v  \sim -45$ km/s  at the center of the Hercules stream or $u$-anomaly
and $v \sim -30$ km/s at the edge or boundary of the stream.

The stream itself is quite wide along the $u$ direction  (over 100 km/s wide)
and comparatively quite narrow in the $v$ direction (40 km/s wide).
In fact the mean $u$ value of the stream is much less than its $u$ width.  
If we use a width in $u$ of $35 \pm 50$km/s then
we estimate a range for $-0.20 \lesssim  dr \lesssim -0.07$ 
in the stream.

In Figure 2 we see the $x_1(2)$ OLR orbits appear
at $dr \sim 0.1$ and not at $dr\sim -0.15$ which is
the location of the Hercules stream or $u$-anomaly.
We could adjust the pattern speed of the bar to move the
resonance over, in fact with $\Omega_s = 1.95$, we can
match the location of the $u$-anomaly.
However before we do this we should consider the approximations
made in our analysis.
In the full Hamiltonian  (Eqn.1) there is a term
proportional to $I_1 I_2$ which, had we kept it in our
analysis, would have caused an additional term 
${b \over a } {I_2 \over \kappa_c r_c^2}$
added to $\bar\delta$ in Eqn.(26).
In dropping the extra terms in the Hamiltonian
we have assumed that $I_2$ is  small.
Because $b$ is the same sign sign as $a$,
$\delta$ should be larger than we have calculated. This
would have the affect of shifting the location of the resonance
to more negative $dr$ or larger distances from the solar neighborhood.

The maps shown in Figures 2--4 do
show the likely morphology of phase space near the OLR.
However we should not use them to constrain the bar
pattern speed at a level better than $25\%$
unless we have taken into account this correction.
We estimate the value of the uncertainty from
the discrepancy between the pattern speed that
we estimate and that constrained by \citep{dehnen99}.

To find out if stars in orbits associated with the $x_1(2)$
family can reach the solar neighborhood, 
we also need to estimate the size of our action
variable for the $u$-anomaly.
If we assume that $u\sim 0$ then the epicyclic amplitude 
is set by the condition that we are at $r_0$.
In other words the maximum value in $r-r_c$ is approximately
$r_0 dr $.  In terms of the coordinate system
of our Poincare maps, a star which
reaches the solar neighborhood has $s \sim dr$.  

In Figure 2, for $dr \sim -0.10$  the quasiperiodic regions
are of size $s \sim 0.15$ which exceeds the value of $dr$.
This is large enough that
stars in these orbits would reach the solar neighborhood.
For $dr \lesssim  0.093$ no $x_1(2)$ orbits exist and no stellar orbits
aligned perpendicular to the bar will reach the solar neighborhood.
Because each value of $dr$ is associated with a $v$ value in
the solar neighborhood velocity distribution this implies
that above a particular value of $v$ no orbits can
be in the quasiperiodic orbit region surrounding
the $x_1(2)$ periodic orbits.
Our model predicts that the $u$-anomaly should
have a sharp boundary in $v$ 
in the solar neighborhood velocity distribution, as observed.
A model lacking spiral structure would predict this
sharp edge as well since it is determined by 
the value of $\delta$ at which the resonance bifurcates (see Figure 1).

In Figure 2 we see that past a certain value of $dr$, corresponding
to a value of $v$, the quasiperiodic orbits are more
likely to be oscillating with the spiral structure than the bar.
This implies that there should be an upper boundary in $v$
to the Hercules stream or $u$-anomaly.

We see in Figure 2 that the quasiperiodic orbits oriented
perpendicular to the bar are bounded by chaotic regions.
This implies that there is a limit in the extent of the
epicyclic amplitude of stars associated with this quasiperiodic
region.   In the solar neighborhood, this implies that there
would be a maximum value of $u$ for orbits in the $u$-anomaly.
Since stars in the chaotic regions can rapidly achieve very different
epicyclic amplitudes, the boundaries of the $u$-anomaly 
are probably set by the extent of the quasiperiodic orbits
associated with the $x_1(2)$ orbits.

\subsection{Changing the spiral pattern speed}

According to \cite{cont88,panos}, 2-armed spiral structure
should only extend between its 2:1 and 4:1 ILRs.
These works showed that 2-armed spiral structure
was not supported by the stellar orbits
past the 4:1 ILR.
At first we consider pattern speeds that are faster,
and so with the solar neighborhood nearer the 2:1 ILR.
The result of integrating Eqn.(\ref{ham_integrate})
with a faster pattern, $\Omega_s = 0.4$, in units of the
solar angular rotation rate $\Omega_0$ is shown
in Figure 3.  This pattern speed would be consistent with a 2-armed
spiral pattern near its 2:1 ILR.

In Figure 3 we see large scale chaotic regions  that completely
disrupt the spiral structure and the orbits associated with
the bar OLR.  
Because $\bar\nu$ is smaller in this integration than
that shown in Figure 2, the resonances are closer together 
and so more fully overlapped.
The scale of the chaotic regions can be reduced by decreasing
the strength of the bar and spiral perturbations or
the values of $\bar\epsilon$, $\bar\beta$.
However, decreasing both $\bar\epsilon$ and $\bar\beta$ by a factor
of two does not restore the quasiperiodic island about the $x_1(2)$ 
family of periodic orbits.
The scale of the chaotic
zone is more sensitive to the extent the resonances overlap 
or to $\bar\nu$ which
is set by the spiral pattern speed, than to the size of
the perturbations.
Since we have adopted realistic values 
for the bar and spiral arm strength
we can use this sensitivity to place an approximate limit
on the pattern speeds of spiral structure that passes 
through the bar OLR.
We find that the $x_1(2)$ are completely disrupted for
$\Omega_s \lesssim 0.45$ implying that 2-armed
spiral structure passing through the OLR is likely to be
rotating faster than this.

In Figure 4 we have integrated a spiral pattern at a faster
pattern speed of $\Omega_s=0.85$ for $m=4$.  This would
correspond to an 4-armed spiral pattern.  The situation is
similar to that seen in Figure 2. 
The OLR bar disrupts the spiral pattern on the outer
side of the resonance.  The $x_1(2)$ orbit family appears
abruptly at $dr \sim0.1$,  and the quasiperiodic island
oscillates with the bar until $dr \sim 0.12$ where 
it more strongly supports the spiral structure.

If we reduce the pattern speed of 4-armed spiral structure
to $\Omega_s < 0.75$ phase space looks like that
shown in Figure 3.  Both the $x_1(2)$ orbit family and quasiperiodic
orbits that support the spiral structure are destroyed.
This implies that 4-armed spiral structure faster
than $\Omega_s = 0.75$ cannot pass through the bar's OLR.


\section{Summary and Discussion}

In this paper we consider the dynamics of stars that
are affected by perturbations from both spiral structure
and the Milky Way bar.
We construct a simple one-dimensional
Hamiltonian model for the strongest resonances
in the epicyclic action and angle variables.

This Hamiltonian is time dependent because the bar and spiral
structure in general have different pattern speeds,
and resembles a forced pendulum.
To address the issue of orbit stability and characterize
the nature of phase space in this system we 
numerically integrate this Hamiltonian.
By plotting points only at
the period which separates the two resonant perturbations we construct
Poincare maps which illustrate where
in phase space there are area filling or chaotic orbits instead of 
quasiperiodic orbits.

The Liapunov time in the chaotic regions is of order
the period which separates two resonances. 
Over much of the resonance this timescale is
of order a few times the rotation period.
Consequently stars which are located
in chaotic regions can move across the 
region in a few rotation periods.   This implies that they
can achieve large epicyclic amplitudes and will not remain
in any coherent small region in phase space.
Streams seen in the local velocity distribution are unlikely
to be located in large chaotic regions.
In contrast,
particles which are located in quasiperiodic regions
should maintain their epicyclic amplitudes for much longer times.
Our model predicts that the chaotic regions can form large bands.
This is a qualitatively different picture than the narrow regions
spanning a range of Liapunov times
illustrated by \citet{fux} who considered solely perturbations from
the bar.  

We find that larger spiral pattern speeds $\Omega_s$ cause
less overlap in the two resonances and so smaller
chaotic zones.
Spiral structure at a pattern speed
which puts the solar circle is near the 2:1 ILR produces extremely large
bands of chaos near the 2:1 OLR with the bar.
This disrupts the spiral pattern and destroys 
orbits perpendicular to the bar near the OLR.
It is unlikely that 2-armed spiral structure 
at patterns faster than $\Omega_s = 0.45$ extends into the OLR.
Likewise, it is unlikely that 4-armed spiral 
faster than $\Omega_s = 0.85$ extends into the OLR.
It is possible that the OLR forms a boundary
between spiral structures at different pattern speeds.

2-armed and 4-armed spiral structure at slower pattern speeds 
are much less disruptive.  For a 2-armed pattern with $\Omega_s = 0.6$, 
chaotic regions appear near the OLR
which disrupt the spiral pattern only across a narrow region.
Orbits oriented perpendicular to the bar appear at asymmetric drift velocities
and epicyclic amplitudes consistent with existence of quasiperiodic 
or stable orbits in the local velocity distribution 
at the Hercules stream or $u$-anomaly.
The boundaries of the $u$-anomaly are set by the extent
of the quasiperiodic orbits associated with the family
of orbits perpendicular to the bar.


In this paper we have considered the combined problem
of a resonance from spiral structure and
from a Galactic bar.  Notably the solar neighborhood 
contains a wealth of structure at smaller velocities 
than the $u$-anomaly \citep{dehnen98}.
Much of this structure contains old stars and so
is likely to be caused by spiral structure.
Since there is more than one clump in the distribution it is likely
that the solar neighborhood is influenced by spiral structure
at more than one pattern speed.  Hence we expect complicated
phenomena such as that discussed here.
In future work we will concentrate on the interplay
between different modes of spiral structure.

\subsection{Caveats}


Our simple Hamiltonian model neglects the coupling between
$I_1$, the epicyclic action variable and $I_2$ which is 
conjugate to the azimuthal angle and related to the angular
momentum.   By neglecting this coupling, we have assumed
that the particle angular momentum is nearly conserved.
Notably in Eqn.~1 the coefficient $b$ is of the same order
as $a$, implying that neglecting the $b$ term is 
only a good approximation when $I_2$ is small.   

Nevertheless taking the limit of small $I_2$ is reasonable
first approach and we feel
that our simplified model is likely to qualitatively illustrate
what happens when multiple resonances are present.
We expect that by considering the entire 2-dimensional
system other phenomena may be exhibited,
for example particles integrated numerically in
a system with both spiral structure and a bar 
can drift in both angular momentum and energy.





\begin{figure*}
\vspace{18.0cm}
\figurenum{2}
\includegraphics{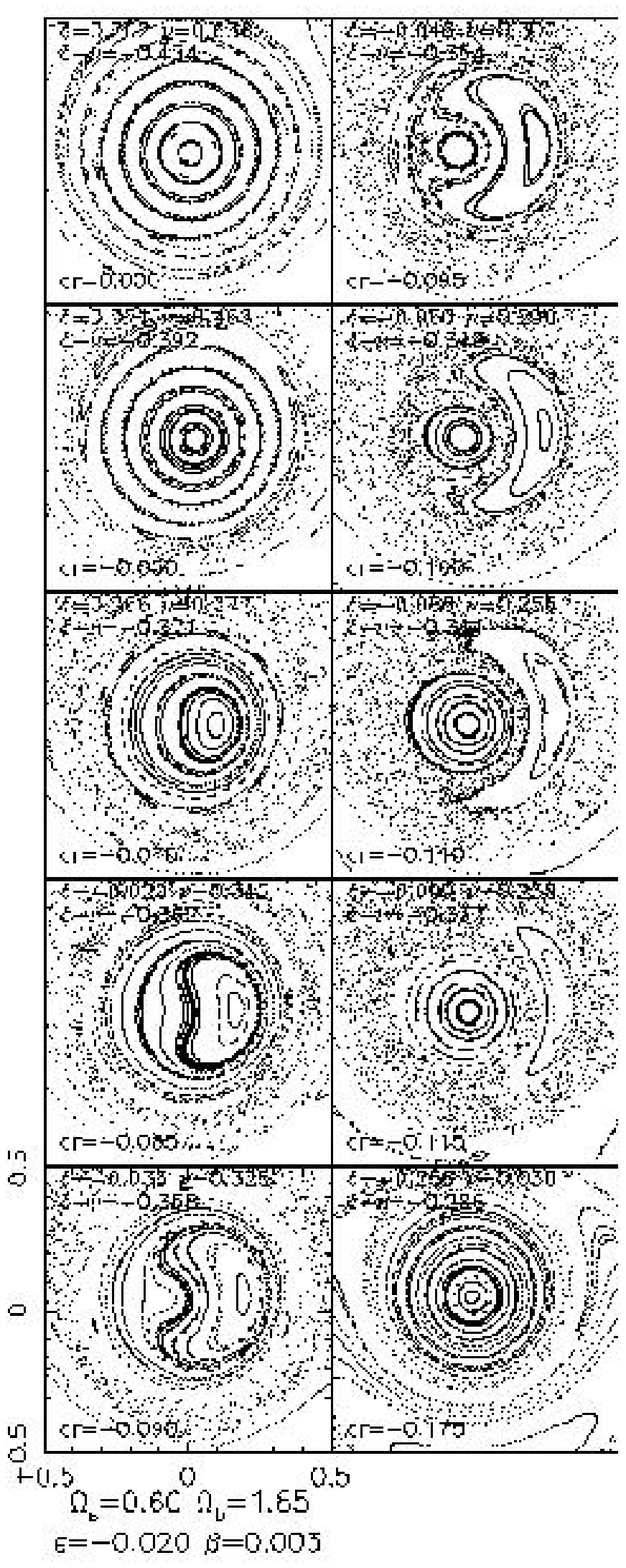}
\includegraphics{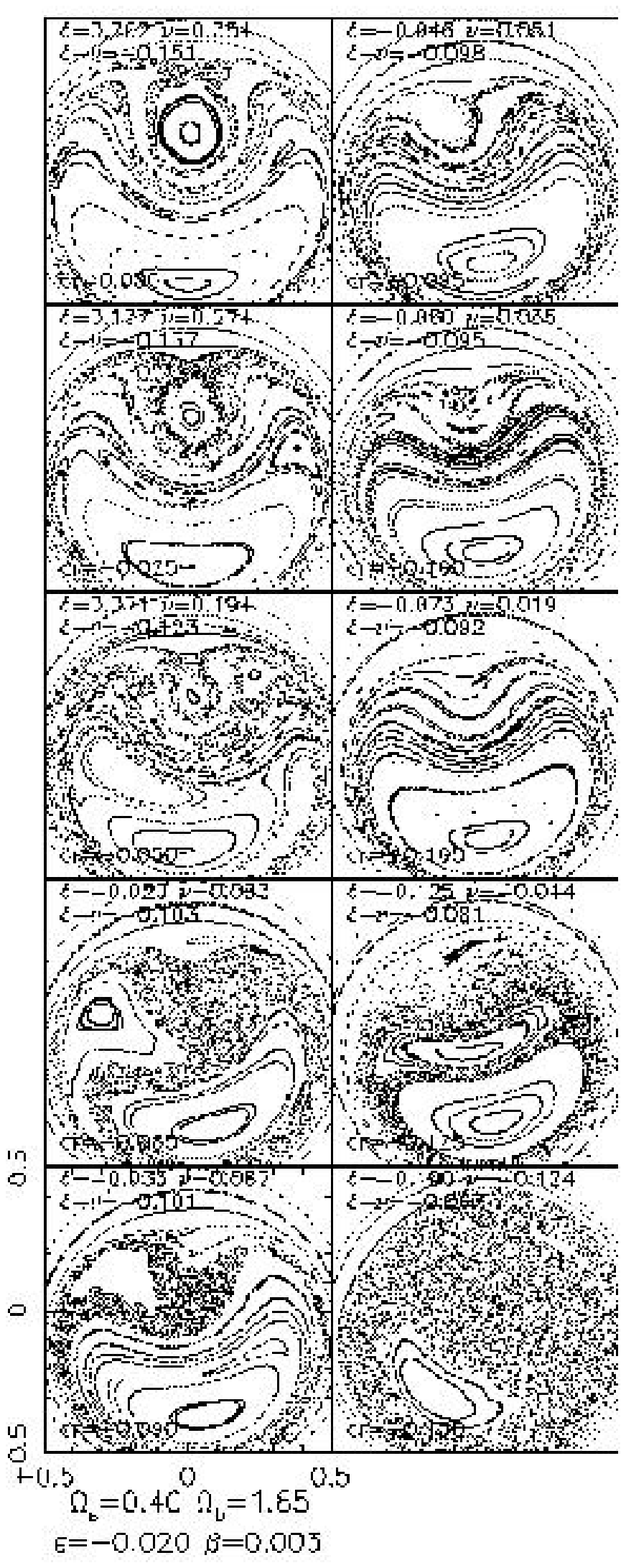}
{
\begin{multicols}{2}
\footnotesize
Fig.~2. -- Poincare maps made by integrating the Hamiltonian of Eqn(\ref{ham_integrate})
with $m=2$ and with a time step of $2 \pi \over \nu$.
The panels are at different values of $dr$.  The top left panel
has the smallest value of $dr$ corresponding to a location
near the solar neighborhood.  The lower and righter most panel
has the most negative value of $dr$ corresponding to a radius
closer to the Galactic center.
All panels have the same values of $\bar\epsilon$,
describing the strength of spiral structure, $\Omega_s$
the pattern speed of the spiral structure in units of the
solar neighborhood angular rotation rate ($\Omega_0$), 
$\bar\beta$, describing the strength of the bar, 
and $\Omega_b$, the pattern speed of the bar in units of $\Omega_0$.
Phase space, as illustrated by the structure in these maps,
has two types of orbits;   curved linear structures, corresponding
quasiperiodic orbits, and area filling orbits  
corresponding to chaotic regions. 
The radial distance on these plots is the same as the 
radial epicyclic amplitude in units of $r_0$, or the distance
the orbit reaches from $r_c$.
A dot is placed at the origin for reference.
The spiral pattern speed considered here would result from a 2-armed spiral
pattern with the solar neighborhood just within the 4:1 ILR.
%

Fig.~3. -- Similar to Figure 2 but we have changed the pattern speed
of the spiral structure to $\Omega_s = 0.4$.
This pattern speed could result from a 2-armed spiral
pattern with the solar neighborhood fairly near the 2:1 ILR.
Since $\nu$ is smaller the resonances are more fully
overlapped than is the case for Figure 2.
Both the spiral structure and resonant orbits at the OLR
are disrupted.
A resonant island of $x_1(2)$ orbits do not exist 
for $\Omega_s \lesssim 0.45$.   The stability
of this orbit family is more sensitive on $\nu$ or the
spiral pattern pattern speed than the strength of the spiral
and bar perturbations.
\bigskip 
~~~~~~~~~~~~~~~~~~~~~~~~~~~~~~~~~~~~~~~~~~~~~~~~~~~~~~~~~~~~~~~~~~~~
\bigskip 
~~~~~~~~~~~~~~~~~~~~~~~~~~~~~~~~~~~~~~~~~~~~~~~~~~~~~~~~~~~~~~~~~~~~
\bigskip 
~~~~~~~~~~~~~~~~~~~~~~~~~~~~~~~~~~~~~~~~~~~~~~~~~~~~~~~~~~~~~~~~~~~~
\bigskip 
~~~~~~~~~~~~~~~~~~~~~~~~~~~~~~~~~~~~~~~~~~~~~~~~~~~~~~~~~~~~~~~~~~~~

\bigskip 
~~~~~~~~~~~~~~~~~~~~~~~~~~~~~~~~~~~~~~~~~~~~~~~~~~~~~~~~~~~~~~~~~~~~
\bigskip 
\end{multicols}
}
\end{figure*}

\begin{figure*}
\vspace{18.0cm}
\figurenum{4}
\includegraphics{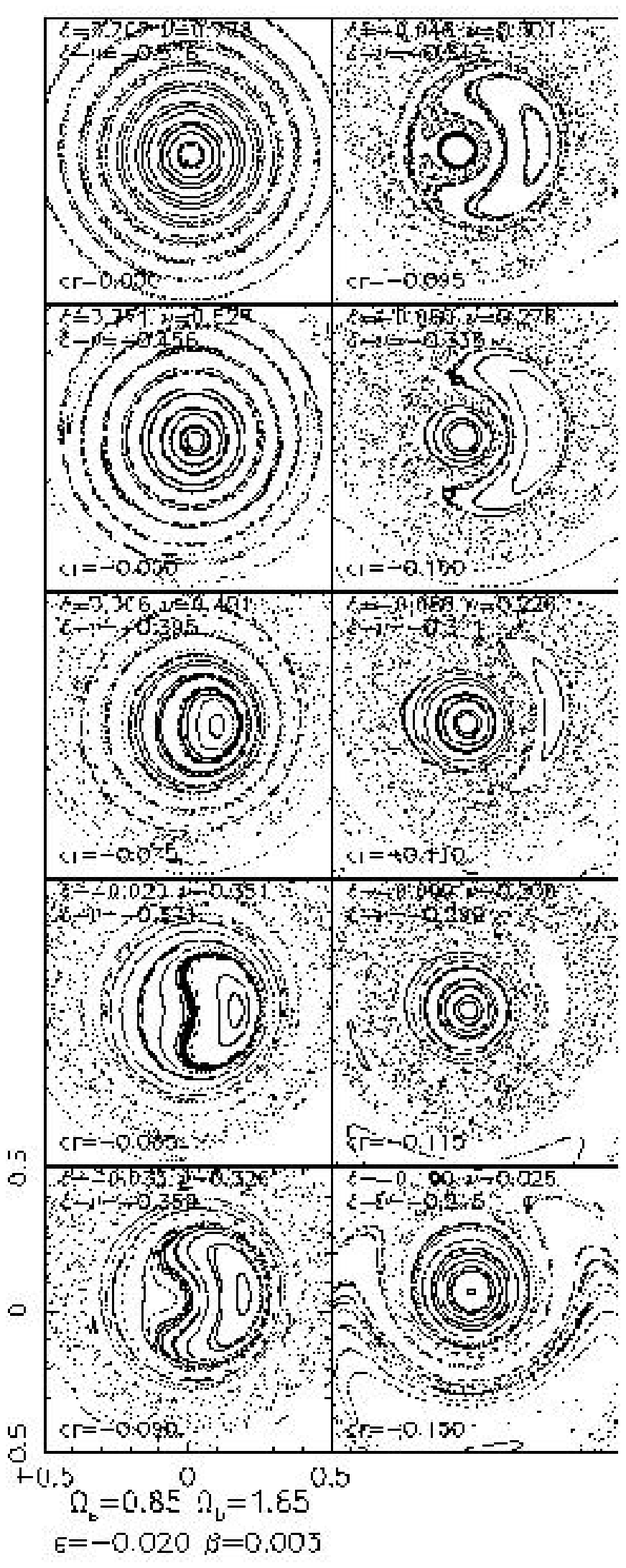}
\caption[]{
Similar to Figure 2 but we have changed the pattern speed
of the spiral structure to $\Omega_s = 0.85$, with
the solar neighborhood outside the 4:1 ILR.
For this simulation $m=4$.
This system would correspond to a four armed spiral pattern.
Phase space as illustrated by this map is similar to that shown
in Figure 2.
}
\end{figure*}


\acknowledgments

This work could not have been carried out without helpful
discussions with Larry Helfer and Don Garnett.
%
%
%
A.~C.~Q.~gratefully thanks the Technion for hospitality and support 
during the fall of 2001.



{}


\begin{thebibliography}{}

\bibitem[Binney \& Tremaine(1987)]{B+T}
Binney, J., \& Tremaine, S.\ 1987,  Galactic Dynamics,
Princeton University Press, Princeton, NJ

\bibitem[Borderies \& Goldreich(1984)]{borderies}
Borderies, N., \& Goldreich, P.~1984, Celestial Mechanics, 32, 127

\bibitem[Contopoulos(1975)]{cont75}
Contopoulos, G.~1975, ApJ, 201, 566

\bibitem[Contopoulos(1988)]{cont88}
Contopoulos, G.~1988, A\&A, 201, 44 

\bibitem[Dehnen(1998)]{dehnen98}
Dehnen, W.~1998, AJ, 115, 2384

\bibitem[Dehnen(1999)]{dehnen99}
Dehnen, W.~1999, ApJ, 524, L34

\bibitem[Dehnen(2000)]{dehnen20}
Dehnen, W.~2000, AJ, 119, 800

\bibitem[Drimmel \& Spergel(2001)]{drimmel}
Drimmel, R., \& Spergel, D.~N.~2001, ApJ, 556, 181

\bibitem[Dwek et al.(1995)]{dwek}
Dwek, E.~et al.~1995, ApJ, 445, 716

\bibitem[Fux(2001)]{fux}
Fux, R.~2001, A\&A 373, 511 

\bibitem[Gerhard(2002)]{gerhard}
Gerhard, O.~2002, 
to appear in: The Dynamics, Structure and History of Galaxies, 
a conference in honor of Ken Freeman, eds. G.S. Da Costa \&
E.M.~Sadler, ASP Conference Series, (astro-ph/0203109)

\bibitem[Holman \& Murray(1996)]{holman}
Holman, M.~J., \& Murray, N.~W.\ 1996, AJ, 112, 127

\bibitem[Holmberg \& Flynn(2000)]{holmberg}
Holmberg, J. \& Flynn, C.~2000, 313, 209



\bibitem[Murray \& Dermott(1999)]{SolarSystemDynamics}
Murray, C.~D. \&  Dermott, S.~F.~1999, Solar System Dynamics,
Cambridge University Press, Cambridge


\bibitem[Patsis \& Kaufmann(1999)]{panos}
Patsis, P.~A., \& Kaufmann, D.~E.~1999, A\&A, 352, 469

\bibitem[Quillen(2002)]{vertical}
Quillen, A.~C.~(2002), submitted to AJ, astro-ph/0203170

\bibitem[Raboud et al.(1998)]{raboud}
Raboud, D., Grenon, M., Martinet, L., Fux, R., \& Udry, S.~1998, A\&A, 335, L61

\bibitem[Skuljan et al.(1999)]{skuljan}
Skuljan, J., Hearnshaw, J. B., \& Cottrell, P.~L.~1999, MNRAS,
308, 731


\bibitem[Taylor \& Cordes(1993)]{taylor}
Taylor, J.~H., \& Cordes, J.M.~1993, ApJ 411, 675

\bibitem[Vall\'ee(1995)]{vallee}
Vall\'ee J.~P.\ 1995, ApJ, 454, 119

\bibitem[Weinberg(1994)]{weinberg}
Weinberg, M. D.~1994, ApJ, 420, 597


\end{thebibliography}
\end{document}